# The laser synthesis of linear carbon chains

Stella Kutrovskaya (1 and 2 and 3) , Anton Osipov (1), Alexey Povolotskiy (4), Vlad Samyshkin (1), Alina Karabchevskii (5), Alexey Kavokin (2 and 3 and 6 and 7) and Alexey Kucherik (1)

((1)Department of Physics and applied mathematics, Stoletov Vladimir State University, 600000 Gor'kii street, Vladimir, Russia, (2) Institute of Natural Sciences, Westlake University, No.18, Shilongshan Road, Cloud Town, Xihu District, Hangzhou, China, (3) Russian Quantum Center, 143025 Novaya street, Skolkovo, Moscow Region, Russia, (4) Institute of Chemistry, St. Petersburg State University, 198504, Ulianovskaya str. 5, St. Petersburg, Russia, (5) Electrooptical Engineering Unit and Ilse Katz Institute for Nanoscale Science & Technology, Ben-Gurion University, Beer-Sheva 84105, Israel, (6) Spin Optics Laboratory, St. Petersburg State University, 198504 Ulyanovskaya street 1, St. Petersburg, Russia, (7) CNR-SPIN, Viale del Politecnico 1, I-00133, Rome, Italy)

**We synthesize macroscopically long linear carbon chains (carbynes) in a colloidal solution, then deposit them on a surface. The method is based on the formation of carbon threads by laser ablation in a colloid accompanied by the stabilization of resulting linear carbon chains by golden nanoparticles. We observe the signatures of the polyyne allotrope of carbyne in the photoluminescence and Raman spectra of the solutions. We deposit the synthesized carbyne threads on fused quartz glass substrates. The transmission electron microscopy images of the deposited threads with golden nanoparticles attached to their ends demonstrate the straight linear parts of the monoatomic carbon chains varying in length between 8 and 24 carbon atoms. This method paves the way to fabrication of an ultimate one-dimensional crystal: a monoatomic carbon wire.**

A variety of low-dimensional crystals based on carbon is in the focus of attention of physical and chemical research communities for several decades. Nanodiamonds, fullerens, carbon nanotubes, graphene bring a unique new physics and promise breath-taking applications in nano-electronics and photonics [1]. One of the most challenging goals for the nanofabrication technologies is the realization of ultimate one-dimensional crystals: monoatomic chains of carbon or carbynes. Traces of two stable allotropes of carbyne (polyyne and cumulene) have been found in nature: in meteorite craters, interstellar dust, natural graphite, diamond mines [2-4]. A high chemical reactivity of carbyne and its low stability at room temperature and atmospheric pressure make it difficult to extract free standing carbon chains from these natural sources. Moreover, multiple attempts to synthesize carbyne artificially, yielded modest or no success so far [5]. Till now, stable freestanding samples of straight carbon chains have not been realised, to the best of our knowledge. Their synthesis appears to be a formidable challenge as, in general, infinite one-dimensional atomic chains are unstable in vacuum. Fluctuations prevent formation of ideal one-dimensional crystals, according to the Landau theorem [6]. In this context, many works have been devoted to the artificial stabilization of carbyne with use of heavy anchor atomic groups [7], confinement of carbon chains in double-wall carbon nanotubes [8], their pinning to metallic surfaces etc. The synthesis of free-standing carbon chains remains one of the Holy Grails of nano-physics and nano-chemistry. If realized these one-dimensional crystals would exhibit unique mechanical, optical and electronic properties [9]. According to the recent theoretical works [10] carbyne would be the most robust of known crystals. In a two-dimensional material made of carbine chains the interatomic distances are predicted to be 0.128 nm along the chain, and 0.295 nm between neighbouring chains, that is significantly less than in graphite where the spacing between neighbouring atoms and the spacing between atomic planes are of 0.142 nm and 0.335 nm, respectively. The predicted Young modulus of carbyne is one order of magnitude larger than that of diamond [11]. The electrical conductivity and the magnetic properties of carbynes may be dramatically varied by stretching and twisting the chain [12]. The strain induced switching of conductivity in carbyne has been addressed recently [13]. The isolated monoatomic chains would demonstrate peculiar spin-dependent quantum transport effects [14]. Moreover, carbynes are predicted to exhibit the superconductivity at high temperatures [15] In addition, carbynes are expected to exhibite unusual optical properties, in particular a giant non-linear optical response [16] Clearly, the synthesis of carbyne with a variable types of bonds is of a very significant interest from the fundamental point of view and for the development of new integrated circuits composed of advanced, next-generation hybrid electronic and photonic devices.

Here we report on the synthesis of stable macroscopically long carbon chains in a colloidal solution by the laser ablation technique. The stabilisation of carbine is achieved due to the stretching of carbon chains between golden nanoparticles. We obtain the polyyne allotropic form of carbyne, which is characterised by specific discrete vibron modes clearly seen in the Raman spectra. When deposited on a substrate, the stabilized chains demonstrate straight parts of the length significantly exceeding the theoretical length limit for a free stable monoatomic carbon chain. The golden nanoparticles attached to the ends of each chain may be used as contacts allowing for the efficient electronic injection and quantum conductivity measurements. This study paves the way to realisation of integrated circuits based on ultimate one-dimensional crystals: the carbynes.

**The synthesis of carbynes in a liquid**

The method of synthesis of carbon chains employed here is described in detail in Methods and the Supplementary Material, see also [17]. We use the double stage laser ablation process. For the stabilisation of long linear chains in the solution between the first and the second stages of the laser action we add golden nanoparticles with an average size of 10 nm (see. Fig. 1). Carbon atoms at the ends of each chain possess uncompensated electrons, while the surfaces of golden nanoparticles are characterized by excess negative charges. For this reason, golden nanoparticles are capable of efficiently catalysing the growth of carbon by pulling the end carbon atoms into the crystal lattice of a metal [18]. The activation energy of an electron in gold is sufficiently low, that simplifies the formation of stable metal-carbon bonds. The length of carbon chains increases due to the association of several linear carbon fragments to one attached to the gold nanoparticle. The process of association comes to its end once another golden nanoparticle traps the carbon atom at the opposite end of the chain. Those carbon chains that have only one end fixed at the surface of a golden nanoparticle and the other end free usually undergo distortions forming a carbon cluster attached to the nanoparticle. Free carbon atoms either diffuse out of the heated by laser volume, or react with oxygen and hydrogen and evaporate. Carbon clusters that are not associated with golden nanoparticles dissociate and form linear chains again, once they come under laser irradiation. This way they supply building blocks to the growing stable carbon chains linked to the golden anchors. It is important to note that the golden anchors stabilize the carbon chains by preventing their vibration-induced decomposition into shorter components, folding and bending. Importantly, the color of the solution and the size distribution of metal nanoparticles are not affected by the laser irradiation process that certifies of the form preservation of metallic nanoparticles associated with carbon chains.

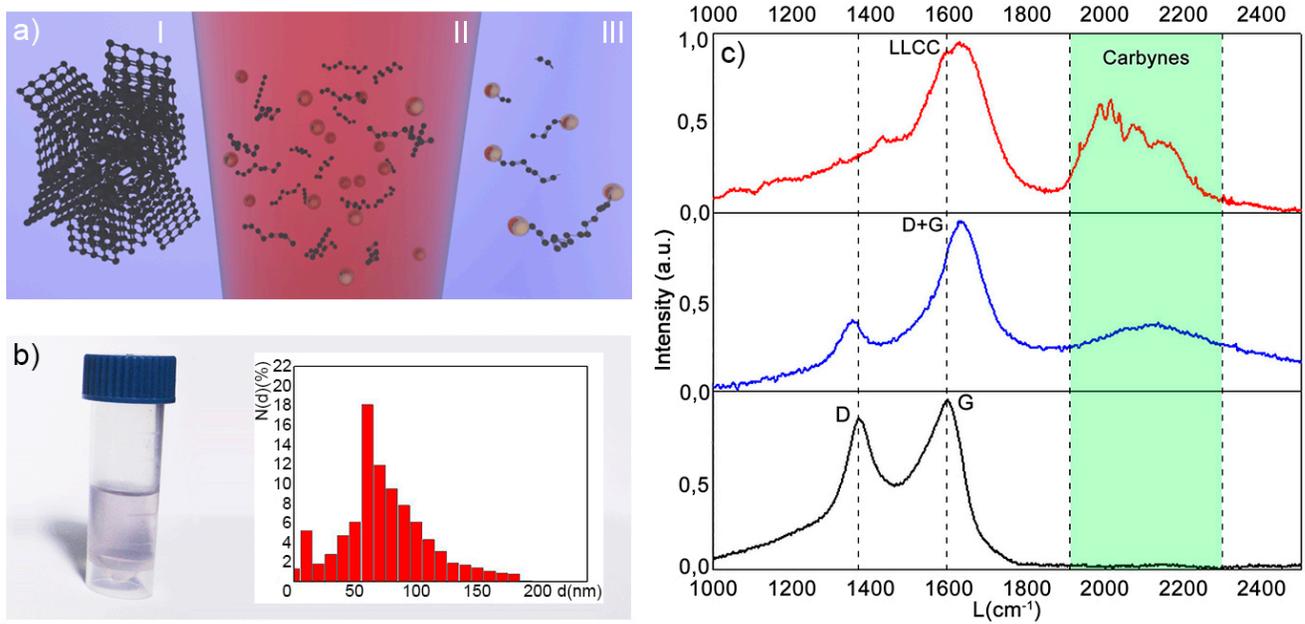

**Fig. 1 (a) Carbyne synthesis and Raman characterization.** Schematically illustrates the mechanism of self-assembly of carbon chains in a colloidal solution under an effect of laser irradiation, (I) showing the amorphous shungite particle, (II) showing the carbon chains and metallic nanoparticles under laser irradiation, (III) showing the monochain carbyne stabilized by golden nanoparticles. **(b)** shows the photo image of a colloidal system carbyne-gold at the end of the synthesis (left) and the diagram of colloidal elements size obtained by the dynamical light scattering method (right), **(c)** shows the Raman spectra of the solution at the different stages of the laser ablation process. The bottom (black) curve represents the spectrum of an initial colloidal shungite particles, the middle (blue) curve corresponds to the unstabilized carbon threads in a solution, the upper (red) curve shows the Raman spectra of carbon chains stabilized with gold nanoparticles. The peaks correspond to characteristic vibron modes of the isolated linear chains of various lengths.

Fig. 1b shows the size distribution of the nano-elements in the colloidal solution at the final stage of the laser synthesis obtained by the dynamical light scattering (DLS) method. The peak at 10 nm corresponds to the characteristic initial size of golden nanoparticles. The size distribution of sp-hybridized carbon based elongated nanostructures peaks at 60 nm. A full set of data on the variation of the size distribution histogram at different stages of the synthesis is presented in the Supplementary material. The synthesised nanostructures in a colloidal solution are characterized by the Raman spectra shown in Fig. 1c. The black curve in this figure shows the Raman spectrum of the initial pure carbon phase. It features two pronounced peaks, namely, D (L=1,380 cm$^{-1}$) and G (L=1,580 cm$^{-1}$) corresponding to the initial sample of shungite. The blue Raman curve corresponds to the irradiated carbon particles in a colloidal system. The analysis of the Raman spectra of the resulting solution shows that the carbon chains are formed at the laser intensities not lower than $10^6$ W/cm$^2$ [19]. We observe the degradation of the D-peak in time indicating the deformation of graphene layers and the formation of a D+G peak [20]. The signature of the carbyne crystal structure is very well seen in the spectral range of 1,900-2,300 cm$^{-1}$. The vibrations of the triple carbon bonds (C≡C)$_n$ corresponding to the polyyne carbyne phase contribute to the band of 2,100-2,300 cm$^{-1}$. The broadening of these peaks during the fabrication process resulting in the formation of a wide spectral band of weak intensity is explained by the beginning of a formation process of long linear carbon chains (LLCC) structures of different lengths.

Whilst golden nanoparticles were added to the colloid system the formation of the merged LLCC peak was observed instead of the separately spaced D and G peaks. The intensity of the bands in the range of 1,900–2,300 cm$^{-1}$ strongly varies in this case. The set of local narrow peaks is observed instead of a single wide peak (Fig. 1c, red curve). The association of carbon chains with golden anchors results in a strong variation of magnitudes of the different bands in the spectrum. Certain vibration modes of carbyne are strongly amplified in the presence of golden nanoparticles. The observed dramatic transformation of the spectra also manifests the change of the length of the newly formed carbon chains: vibrations of polyyne bonds for the chains of different lengths amplified by golden nano-antennas give rise to the various narrow intensity peaks. In particular, the peak at ~2,000 cm$^{-1}$ corresponds to the bond length alternation (BLA) mode that can be considered as a signature of the linear carbon chain formation [21]. We observe a set of high intensity peaks in the range of 1,950-2,050 cm$^{-1}$ that correspond to BLA modes of the chains of different lengths. Lower intensity peaks within the 1,950-2,050 cm$^{-1}$ range may be attributed to the chains containing 30-60° kinks separated by 8-16 atom long linear parts [21].

The absorbance spectra of solutions containing the carbon chains (see the panel (a) of fig 2) demonstrate a characteristic carbon absorption band at the wavelength range of 240-350 nm and a band corresponding to the plasmon absorption of gold nanoparticles (near 510 nm ). In the inset to Fig. 2a one can see that the highest magnitude absorption is detected in the spectral range of 240 nm. This band corresponds to the absorption of light by linear carbon chains containing 10 atoms. [22]. We observe, in particular, the local maxima of absorption in the spectral range of 240-450 nm that are characteristic of $C_{10}$ - $C_{26}$ linear chains of carbon atoms. These peaks may be accurately fitted by Gaussian functions.

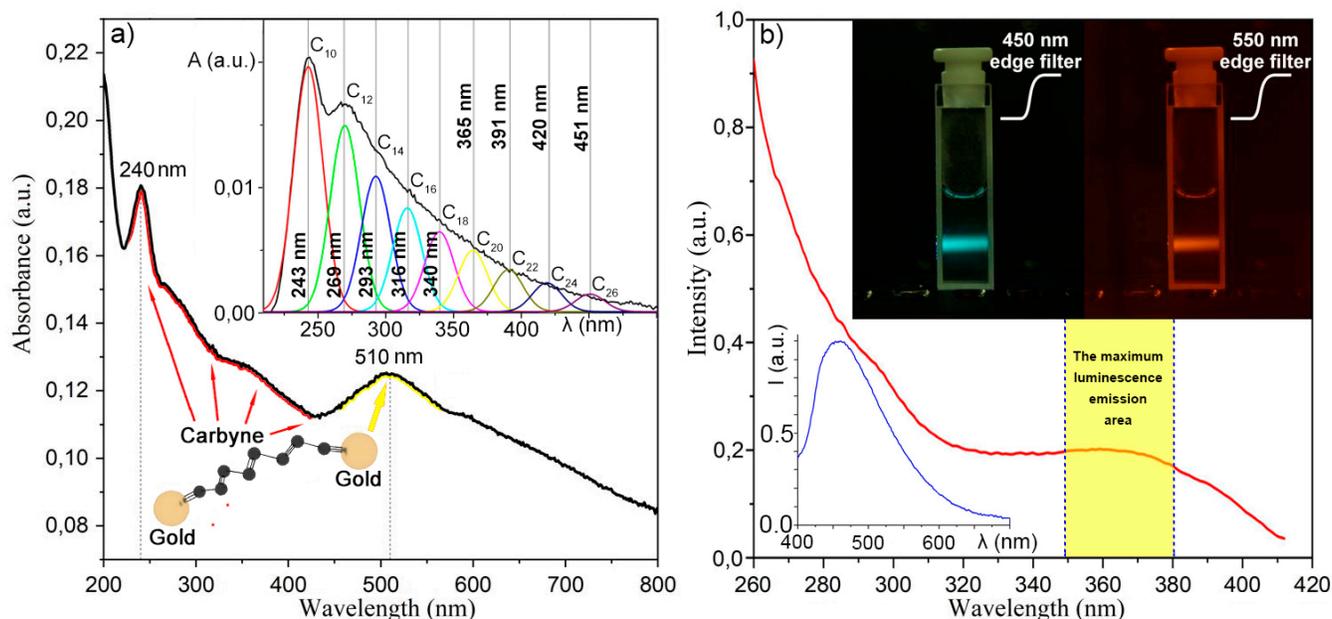

**Fig. 2 Optical investigation of colloidal solution. a)** The absorption spectra of the carbyne chains stabilised by golden ancors in a solution; the inset subtracting the exponential background characteristic of a fundamental absorption band, we reveal a long sequence of resonances going up the wavelength of 450 nm, **b)** The photoluminescence excitation (red line) and photoluminescence (blue line) spectra of the solution. The photoluminescence of carbynes in the solution was excited by the third harmonic of neodymium laser (355 nm) and detected through edge filters, the inset to the right panel showing the photo images of the luminescent solution taken with 450 nm and 550 nm filters. The filters are employed to block the scattered light of the laser.

The further insite into the structural properties of the synthesised carbon chains is provided by the spectra of photoluminescence excitation detected with use of the Horiba Fluorolog-3 spectrofluorimeter with an 8 nm spectral width slit. The detection wavelength was fixed at 442 nm, while the pump wavelength varied in the range of 260 to 430 nm (Fig 2b, red line). The characteristic rise of the emission signal was detected at the excitation wavelengths of 320-430 nm. This is characteristic of the photoluminescence excitation through the transitions between the highest occupied molecular orbital (HOMO) and the lowest unuccupied molecular orbital (LUMO) with a subsequent LUMO-HOMO recombination at 442 nm. The photoluminescence spectrum of the solution (blue line in Fig. 2b) indicates that the HOMO-LUMO splitting in the synthesised carbynes varies in the range of 400-600 nm. Note that the width of the energy gap in linear carbon chains is sensitive to the length of the chain. The photoluminescence quenching in a colloidal system prevents the detection of the resonances apparent in the absorption spectra (Fig. 2a). On the other hand, it provided us with the estimate of LUMO-HOMO splitting of 2.9 eV and larger that is consistent with the predictions by the ab-initio calculation for a carbon chain of 8-16 atoms [23]. The absorption spectra (Fig. 2a) indicate that the most frequently found chains contain 10 atoms while some of the chains achieve the maximum length of 26 atoms.

**Deposition of stabilized carbon chains on a substrate**

Application of carbone chains in nano-electronics requires their deposition on a surface. We have deposited on a fused quartz glass substrate and dried out the droplets of the colloidal solution containing carbon chains stabilised by golden anchors. Fig. 3(a) shows the excitation of photoluminescence detected at the wavelength of 442 nm as well the photoluminescence spectrum excited at 370 nm. The spectra are characteristic for molecular systems: the spectra overlap, and the Levshin mirror symmetry is clearly observed. The Stokes shift induced by non-radiative relaxation processes is about 4400 см$^{-1}$. According to Pan *et al* [23], the emission maxima at 410, 435 and 465 nm correspond to LUMO→HOMO transitions in carbon chains composed by 8, 10 and 12 atoms, respectively. Following the approach of Ref. 23, we have approximated the observed broad photoluminescence band by a set of Gaussian peaks corresponding to the different lengths of the carbon chains. We have assumed the same broadening of all Gaussian peaks to reduce the number of free parameters of the model. This analysis allowed us to plot the distribution function of the lengths of the straight linear parts of the carbon chains deposited on the surface (see Fig. 3d).

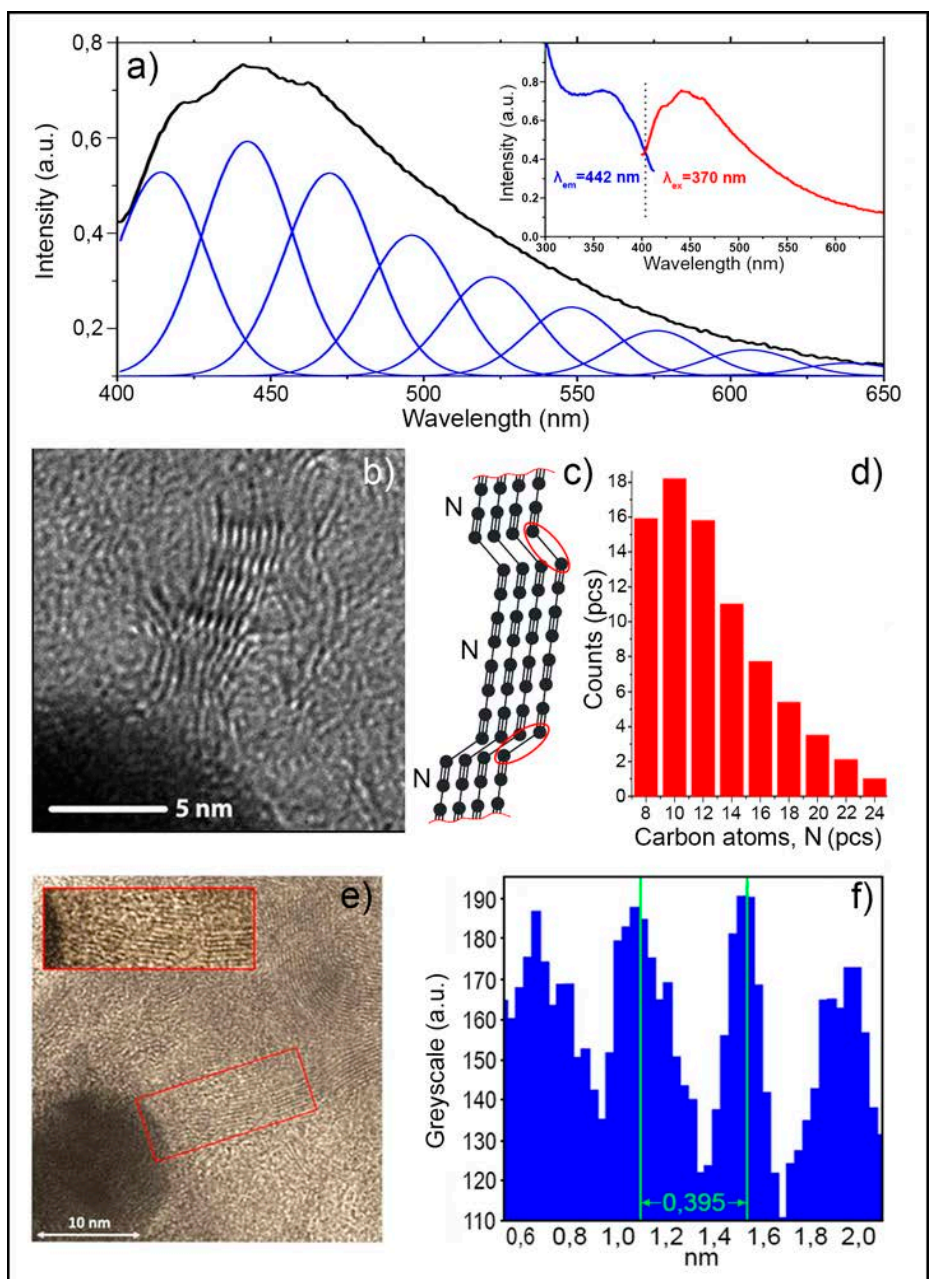

**Fig.3 Deposited chains investigation. (a)** The photoluminescence spectrum of a carbyne film deposited on a fused quartz glass substrate (black line) and its fit by a distribution of Gaussian functions corresponding to individual carbon chains. The inset shows the photoluminescence excitation (blue line) and photoluminescence (red line) spectra of the deposited carbynes. **(b)** TEM image of the studied carbyne film. **(c)** shows schematically the atomic structure of deposited carbon chains **(d)** shows the extracted length distribution of the deposited carbon chains. TEM image **(e)** shows the fragments of a two-dimensional crystal structure formed by parallel carbon chains deposited on the substrate. The period of the structure in the direction perpendicular to the chains is extracted from the profile of the TEM image **(f)** as 0.395 nm.

We have identified the chains containing 8, 10 and 12 atoms by comparing their emission wavelength with the data of [23]. Assuming the linear dependence of the HOMO-LUMO gap on the number of atoms in the chain and the proportionality between the concentration of carbon chains of a given length to the magnitude of the corresponding Gaussian peak, we have described the observed photoluminescence band by the superposition of the emission spectra of the broad distribution of linear carbon chains (Fig. 3d). One can see that the chains containing from 8 to 24 carbon atoms dominate the spectrum. 10 atom straight linear chains are the most frequently found ones. It is very important to note that virtually all the observed linear parts of the deposited carbon chains contain an even number of atoms. This is very apparent on the TEM image (Fig. 3b) and it is confirmed by the Gaussian fit in Fig.3a. We attribute this remarkable parity selection rule to the selective bending of polyyne chains at single electronic bonds (Fig. 3c). Single bonds are "weaker" than triple ones, so that the strain accumulated in a chain whose ends are attached by triple bonds to metallic nanoparticles manifests itself in kinks at some of the single bonds. The number of carbon atoms between any two single-electron bonds is even.

One can clearly see also that the lengths of deposited straight linear parts of carbon chains significantly exceed the theoretical limit of 5-6 atoms for free-standing carbynes. We can confidently distinguish free of Peierls deformation [24] linear parts of carbon chains containing up to 24 atoms. The full lengths of the chains including the bended parts (Fig. 3e) achieve several tens of nanometers, that matches the DSL data of Fig. 1(b). Very interestingly, parallel linear carbon chains form a regular two-dimensional crystal structure. The profile of this 2D carbyne crystal is plotted in Fig. 3f. The period in the direction perpendicular to the wires is found to be 0.395 nm, with a good accuracy. The two-dimensional carbyne crystal phase is obtained by simple drying out of the droplets of a colloidal solution containing metal-stabilised carbon chains deposited on a fused quartz glass substrate. No supplementary stabilisation tools were used in this case.

**Conclusion**

As discussed above, the maximum length of the synthesised carbynes stabilised by golden anchors may be estimated as 60 nanometers. In a solution, the Peierls deformation [24] is compensated by the viscosity of the surrounding medium and the attraction to metallic anchors. When deposited on a surface, the carbon chains necessarily fold and bend. Nevertheless, we are still able to identify the straight linear parts containing from 8 to 24 atoms. This significantly exceeds the theoretical estimate of the length limit for a non-perturbed linear carbon chain that is 5-6 atoms [25]. The strong dependence of the band gap of carbyne on the number of atoms in a chain offers an efficient tool of control over the electronic properties of these ultimate nano-wires by the length selection. The present results pave the way to deposition of macroscopically long carbon wires comparable to those present in the solution. As a promising direction towards the realisation of this goal we consider the electrical stretching of carbines stabilised by golden anchors. We underline at this point that the golden anchors offer several important advantages for practical applications of carbon chains. First, the triple electronic bonds of carbon atoms to gold preferentially stabilise the polyyne allotrope of carbyne. We see the signatures of the polyyne structure in the specific parity selection rule on the length of the chain observed both in the solution and after the substrate deposition. Moreover, the significant presence of the polyyne phase is confirmed by the photoluminescence and Raman data. The selective synthesis of a polyyne allotrope may facilitate applications of carbynes in nano-elecrtonics. Second, golden anchors strongly reduce the negative role of Peierls deformation and they allow for the deposition of record long straight linear carbon chains without any supplementary stabilisation. Third, golden nanoparticles may be used as reliable metallic contacts with well-known electro-physical properties that paves the way to quantum transport experiments and to the on-chip integration of the carbon chains.

**Acknowledgements**

This study was supported by the Ministry of Education and Science of the Russian Federation (state project no. 16.5592.2017/VU), Russian Foundation for Basic Research grants # 16-32-60067, # 17-32-50171 and by the grant of president of Russian Federation by project MK-2842.2017.2. A. Kavokin acknowledges the Saint-Petersburg State University for the research grant 11.34.2.2012.

Raman spectra, absorbance and luminescence spectra were measured at the Center for Optical and Laser Materials Research, Research Park, St. Petersburg State University

**Methods**

For realisation of stabilised carbyne chains we have employed the method of laser fragmentation of colloidal carbon systems that is described in detail in Ref. [17] and in the Supplementary material (available online).

**References**

1. Segawa, Y., Ito, H. & Itami, K. Structurally uniform and atomically precise carbon nanostructures. *NatureRev.Mater.* **1**, 15002 (2016).
2. Goresy, A. E. & Donnay, G. A new allotropic form of carbon from ries crater. *Science* **161**, 363-364 (1968).
3. Wittaker, A. G. Carbon: Occurrence of carbyne forms of carbon in natural graphite. *Carbon* **17**, 21-24 (1978).
4. Chuan, X. Y., Zheng, Z. & Chen, J. Flakes of natural carbyne in a diamond mine. *Carbon* **41**, 1877-1880 (2003).
5. Kroto, H. Carbyne and other myths about carbon. *Chem. World* **7**, 37 (2010).
6. Landau, L. D., Lifshitz, E.M. *Statistical Physics, Third Edition, Part 1: Volume 5 (Course of Theoretical Physics, Volume 5).* (Butterworth-Heinemann:Oxford, 1980)
7. Gibtner, T., Hampel, F., Gisselbrecht, J.-P., Hirsch, A. End-cap stabilized oligoynes: model compounds for the linear sp carbon allotrope carbyne. *Chem. Eur. J.* **8**, 408-432 (2002).
8. Shi, L. *et al.* Confined linear carbon chains as a route to bulk carbyne. *Nat. Mater.* **15**, 634-639 (2016).
9. Khanna, R. *et al.* Formation of carbyne-like materials during low temperature pyrolysis of lignocellulosic biomass: A natural resource of linear sp carbons. *Sci. Rep.* **7**, 16832 (2017).
10. Liu, M., Artyukhov, V. I., .Lee, H., Xu, F., Yakobson, B. I. Carbyne from first principles: chain of C atoms, a nanorod or a nanorope. *ACS Nano* **7**, 10075-10082 (2013).
11. Itzhaki, L., Altus, E., Basch, H., Hoz, S. Harder than diamond: determining the cross-sectional area and Young's modulus of molecular rods. *Angew. Chem.* **117,** 7598-7601 (2005).
12. Cretu, O. *et al.* Electrical conductivity measured in atomic carbon chains. *Nano Lett.* **13**, 3487-3493 (2013).
13. La Torre, A., Botello-Mendez, A., Baaziz, W., Charlier, J.-C. & Banhart F. Strain-induced metal-semiconductor transition observed in atomic carbon chains. *Nat. Commun.* **6**, 6636 (2015).
14. Zanolli, Z, Onida, G, Charlier, J. C. Quantum spin transport in carbon chains. *ACS Nano.* **4**, 5174-5180. (2010).
15. Tangetal Z. K. *et al.* Superconductivityin 4 angstrom single-walled carbon nanotubes. *Science* **292**, 2462 (2001).

16. Ma, C. R., Xiaoa, J. & Yang, G. W. Giant nonlinear optical responses of carbine. *J. Mater. Chem. C* **4**, 4692-4698 (2016).
17. Kucherik, A.O., Arakelian, S.M., Garnov S.V., Kutrovskaya S.V., Nogtev, D.S., Osipov, A.V., Khor'kov, K.S. Two-stage laser-induced synthesis of linear carbon chains. *Quantum electron* **46**(7), 627–633. (2016).
18. Yazyev, O. V., Pasquarello, A. Effect of metal elements in catalytic growth of carbon nanotubes. *Phys.Rev.Lett.* **100**, 156102 (2008).
19. Kucherik, A.O. *et al.* Laser-induced synthesis of metal–carbon materials for implementing surface-enhanced Raman scattering. *Optics and Spectroscopy* **121**, 263-270 (2016).
20. Ravagnan, L. *et al.* Cluster-beam deposition and in situ characterization of carbyne-rich carbon films. *Phys. Rev. Lett.* **89**, 285506 (2002).
21. Buntov, E. A., Zatsepin, A. F., Guseva, M. B., & Ponosov, Y. S. 2D-ordered kinked carbyne chains: DFT modeling and Raman characterization. *Carbon*, **117**, 271-278. (2017).
22. Cataldo, F. Synthesis of polyynes in a submerged electric arc in organic solvents. *Carbon* **42**, 129-142 (2004).
23. Pan B. *et al.* Carbyne with finite length: The one-dimensional *sp*carbon. *Sci. Adv.* **1**, e1500857 (2015).
24. Bianco *et al.* A carbon science perspective in 2018: Current achievements and future challenges. *Carbon* **132**, 785-801 (2018).
25. Saito, R., Dresselhaus, G., Dresselhaus, M. S. *Physical Properties of Carbon Nanotubes* (Imperial College Press: London, 1998).

# The laser synthesis of linear carbon chains (Supplementary materials)

Stella Kutrovskaya , Anton Osipov, Alexey Povolotskiy, Vlad Samyshkin, Alina Karabchevskii, Alexey Kavokin, Alexey Kucherik

**The synthesis of carbynes in a liquid**

We use here the method of synthesis of carbyne chains which consists of two stages, also described in [S1, S2]. At the first stage, as a result of laser ablation of a carbon sample placed in a distilled water, we produce a colloidal solution of disoriented carbon clusters (fig.S1a). We chose scungite as a target. The advantage of this target is in its original structure that is composed of a mixture of various carbon allotropes: natural graphene planes and a nanostructured carbon. The carbon crystal nanostructures are usually encapsulated by the amorphous carbon. Note also that shungite is not susceptible to graphitization. At the second stage, we add the small-size golden nanoparticles with diameters close to 10 nm (fig.S1b and S1c) and irradiate the colloidal system by nanosecond laser pulses (Ytterbium (Yb) fiber laser having a central frequency of 1.06 μm, pulse duration of 100 ns, repetition rate of 20 kHz and pulse energy up to 1 μJ, the time between each pulse about 1μsec), which induce the fragmentation of the carbon component of the mixed colloid. In the conditions of our experiments, the maximum temperature achieved due to the irradiation of the colloid by a single laser pulse does not exceed the carbon sublimation temperature (~5000 K) [S1]. In this case, the energy that is absorbed by carbon 100-nm particles is sufficient to provide their partial fragmentation without destroying the linear interatomic bonds. The laser light doesn't provide any thermal action onto the golden nanoparticles because of their small size as it is demonstrated and discussed in detail in [S3]. In our experiments, the laser pulse energy did not exceed 1 μJ. Consequently, the energy absorbed by the carbon nanoparticles was significantly lower than the electron binding energy in an sp2 hybridized carbon (of the order of 7ev/atom [S4]) that would be required for the partial fragmentation of a graphene plane into the carbon chains [S5, S6].  Finally, the long linear carbon chains are formed by self-assembly in the peripheral volume. A significant loss of the color of the solution is characteristic of the carbyne formation (fig.S1d). Carbon linear chains are stabilized by the effect of the environment in a colloidal solution and due to their stretching by golden nanoparticles attached to the ends of the chains.

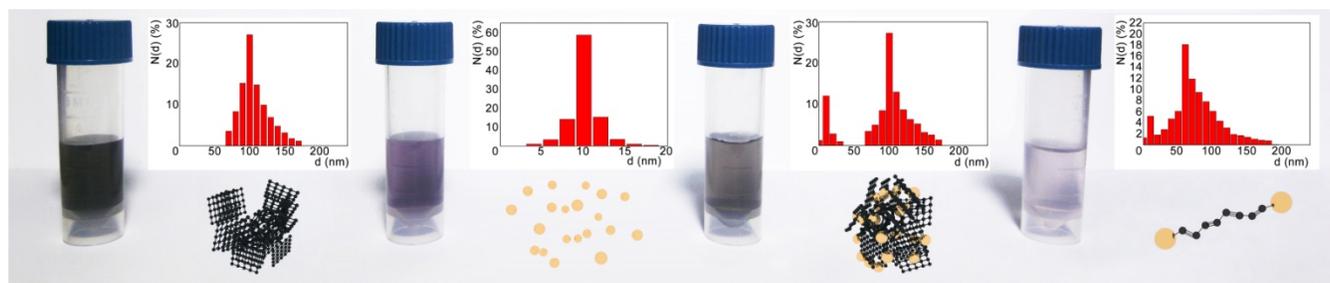

**Figure S1 Colloidal systems with different compositions relevant to the carbyne synthesis procedure we employed.** Left to right: a) the initial carbon system, b) golden nanoparticles, c) the initial mixed solution, d) the mixed system after ns-laser pulse irradiation.

We analyzed the created particles using the dynamic light scattering (DLS) at the Horiba LB-550 and explored the particle's size (exposition time was about 30 sec). These measurements showed how the dispersed phase of liquid systems varied during the synthesis processes. The particle's size distribution of a mixed system certifies of the presence of two non-interacting fractions: the smaller-size Au NPs and the parts of disordered carbon. With the increase of the processing time, the colloidal solution was becaming colorless. The variation of the bimodal histogram (Fig. S1) in time evidenced the interplay of two competitive processes: the fragmentation of carbon nanoarticles under the effect of laser radiation and the self-assembly of carbon atoms into long liner chains formed in the peripheral volumes of the colloid.

**Optical characterisation of the colloidal system**

The spectrophotometric studies of colloidal systems reveal the presence of a set of well-resolved absorption peaks in the spectral range of 200-350nm, that are characteristic of the polyyne chains. Fig. S2. Shows the maxima of absorption that are characteristic of the linear carbon chains of different lengths. Using the data of Refs. [S7, S8] we could associate each of the observed resonances to a carbon chain of a specific lengths, as Table S1 shows.

Table S1

| Peak, nm | Chain length |
|---|---|
| 241 | $C_{10}$ |
| 269, 283 | $C_{12}$ |
| 296 | $C_{14}$ |
| 318, 321 | $C_{16}$ |

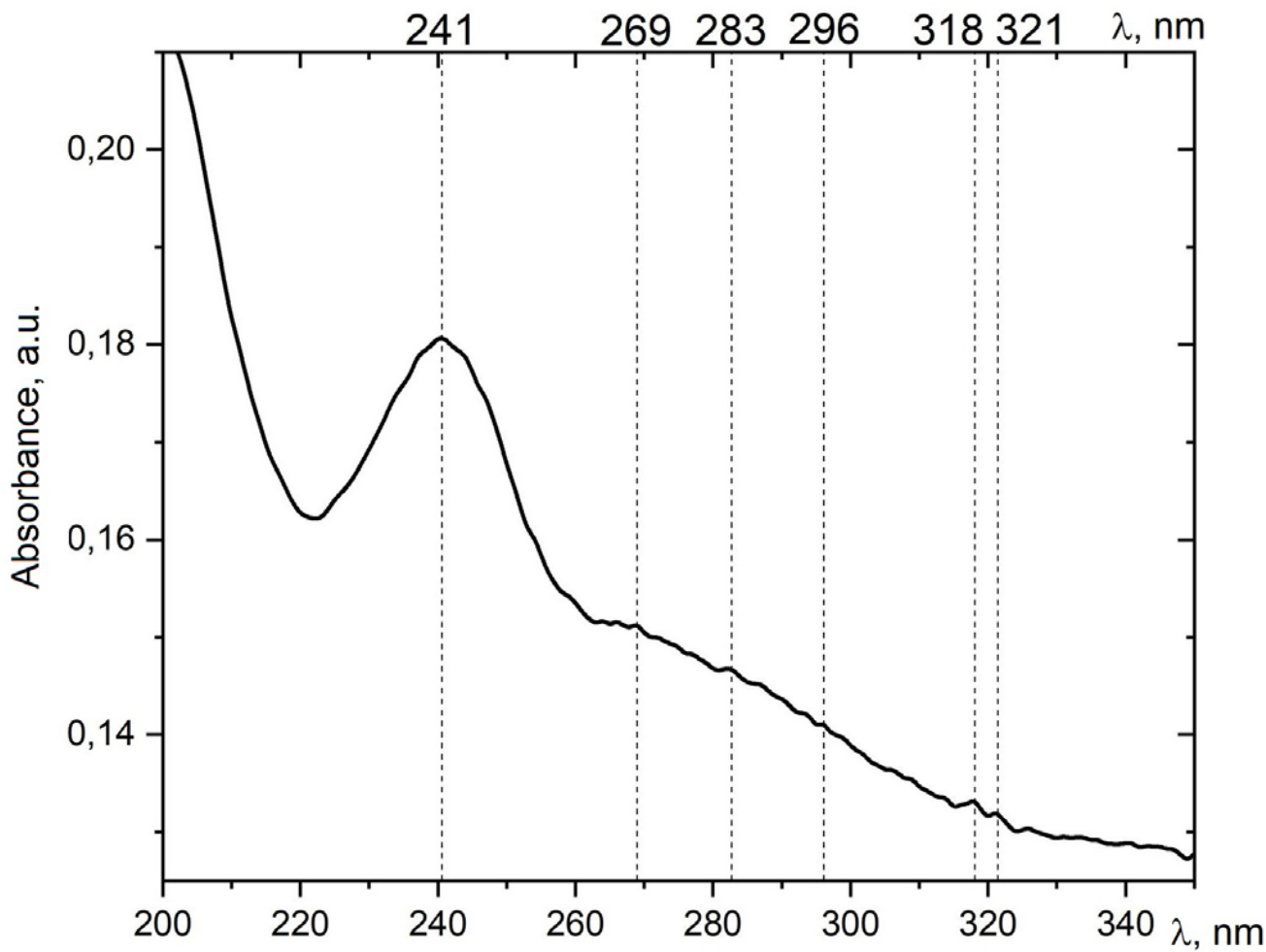

Fig S2. The absorption band exhibiting the sets of resonances characteristic for the polyyne chains of different lengths.

The resolution of the absorption spectrum S2 has been improved by subtracting the background characteristic of the exponential tail of the fundamental absorption band.

**References**

S1. Kucherik, A.O., Arakelian, S.M., Garnov S.V., Kutrovskaya S.V., Nogtev, D.S., Osipov, A.V., Khor'kov, K.S. Two-stage laser-induced synthesis of linear carbon chains. *Quantum electron* **46**(7), 627–633. (2016).

S2. Kucherik, A.O. *et al.* Laser-induced synthesis of metal–carbon materials for implementing surface-enhanced Raman scattering. *Optics and Spectroscopy* **121**, 263-270 (2016

S3. A.Kucherik, Y.Ryabchikov, S.Kutrovskaya, A.Al-Kattan, S.Arakelyan, T.Itina, A.V.Kabashin Cavitation-free continuous-wave laser ablation from a solid target to synthesize low-size-dispersed gold nanoparticles // *European journal of chemical physics and physical chemistry* **18** (9) 1185-1191 (2017)

S4. H.Amara, C. Bichara Modeling the Growth of Single-Wall Carbon Nanotubes// *Top CurrChem (Z)* 375 55 (2017)

S5. C. B. Cannella and N. Goldman Carbyne Fiber Synthesis within Evaporating Metallic Liquid Carbon *J. Phys. Chem. C*, **119** (37), 21605-21611 (2015)
S6. Pan B. *et al.* Carbyne with finite length: The one-dimensional *sp* carbon. *Sci. Adv.* **1**, e1500857 (2015).
S7. N. R. Arutyunyan, V. V. Kononenko, V. M. Gololobov and E. D. Obraztsova Resonant Effects in SERS Spectra of Linear Carbon Chains *Phys. Status Solidi B* 1700254 (2017)
S8. Cataldo, F. Synthesis of polyynes in a submerged electric arc in organic solvents. *Carbon* **42**, 129-142 (2004).